\documentclass[12pt]{book}
\usepackage{plenum}
\begin{document}

\chapter{
MICROSCOPIC UNIVERSALITY IN RANDOM MATRIX MODELS OF QCD
}

\author{
Shinsuke M. Nishigaki
\footnote{Present address: 
Institute for Theoretical Physics, University of California,
Santa Barbara, CA 93106, USA. \ \ 
E-mail: shinsuke@itp.ucsb.edu}
}

\affiliation{
Niels Bohr Institute\newline
University of Copenhagen\newline
Blegdamsvej 17\newline
DK-2100 Copenhagen \O\newline
Denmark}

\vspace{-10cm}
\begin{flushright}
hep-th/9712051\\
NBI-HE-97-46 ~~~\,
\end{flushright}
\vspace{10cm}

\section{INTRODUCTION}
Random Matrix Theory (RMT) has been a useful laboratory
for simulating hamiltonians of statistical systems
with random matrix elements\refnote{\cite{Meh}}, 
due possibly to impurities
scattered around the material.
An unphysical yet simplest choice is to
assume each of matrix elements of the hamiltonian $M$
to be independently derived from the gaussian distribution,
\begin{equation}
Z = \int_{N\times N\ {\rm matrices}}\!\!\!\!\!\!\!\!\!\!
dM\, {\rm e}^{-N\,{\rm tr}\,M^2},
\label{GE}
\end{equation}
to which is associated Wigner's single-banded, semi-circle spectrum.
This macroscopic spectrum is, however, by no means realistic;
physical spectra are far more complicated.
Namely they may even have multi-band structures,
which in the side of RMT correspond to non-gaussian distributions.
Thus, it is only universal quantities insensitive to a 
chosen matrix measure that may be justifiably 
extracted from RMT for physical systems.

Various quantities concerning microscopic spectral correlation
in the bulk of the spectral band have long been believed universal,
as identical results are derived from non-gaussian ensembles
corresponding to classical orthogonal polynomials other
than Hermite\refnote{\cite{Sze}}.
Here the term `microscopic' refers to that the correlation 
is measured in the unit of the mean level spacing, 
which is of order $O(1/N)$ for ensembles of the type (\ref{GE}).
This conjectured universality of the sine kernel
\begin{equation}
K_s(\lambda, \lambda')= 
\frac{\sin\pi(\lambda-\lambda')}{\pi (\lambda-\lambda')},
\label{sine}
\end{equation}
which comprises all spectral correlators
\begin{equation}
\rho_s(\lambda_1,\cdots,\lambda_p)=
\left\langle \prod_{a=1}^p \frac{1}{N} 
{\rm tr}\, \delta (\lambda_a-M)\right\rangle=
\det_{1\leq a,b \leq p} K_s(\lambda_a,\lambda_b),
\end{equation}
is proved\refnote{\cite{BZ,Moo}} for unitary invariant
ensembles with
generic single-trace potentials 
${\rm e}^{-N{\rm tr}V(M)}\!\!.$

RMT has recently extended its range of 
applicability toward QCD\refnote{\cite{SV}}.
There the Dirac operator
$i\,D\!\!\!\!/\ = \gamma_\mu (i\,\partial_\mu + A_\mu)$,
having gauge fields as random elements,
is regarded as a random hamiltonian.
Schematically, the Euclidean QCD partition 
function\footnote{For simplicity, we consider only the 
topologically trivial sector.} 
\begin{eqnarray}
Z_{\rm QCD}&=&
\int [dA_\mu\, d\overline{\psi}_f\, d\psi_f]
\,{\rm e}^{-\frac{1}{g^2}\int {\rm tr} F^2-\int 
\overline{\psi}_f (i\,D\!\!\!\!/ + m_f)\,{\psi}_f
}
\nonumber\\
&=&\int [dA_\mu]
\,{\rm e}^{-\frac{1}{g^2}\int {\rm tr} F^2}\,
\prod_f {\rm Det}
{
\left(
\begin{array}{cc}
m_f & -i\sigma_\mu(A_\mu+i\partial_\mu) \\
i\sigma_\mu(A_\mu+i\partial_\mu) & m_f
\end{array}
\right)}
\end{eqnarray}
is transformed by the change of integration variables 
\begin{equation}
A_\mu\mapsto i\,D\!\!\!\!/_A\equiv M =
{\left(\begin{array}{cc}
0 & W^\dagger \\
W & 0
\end{array}\right)}
\end{equation}
into a RMT
\begin{equation}
Z_{\rm RMT}=\int_{N\times N\,\mbox{\scriptsize matrices}}
\!\!\!\!\!\!\!\!\!\!
d\mu(W)\ \prod_f \det
{
\left(
\begin{array}{cc}
m_f  &  W^\dagger \\
W & m_f
\end{array}
\right)} .
\label{chGUE+m}
\end{equation}
Here $N$ stands for the size of the Dirac operator i.e.
the volume of the spacetime, and
$d\mu(W)$ a measure invariant under
\begin{equation} 
W\rightarrow U\,W\,V^\dagger\ \ 
(U, V \in {\rm U}(N) \mbox{ for $W$ complex}).
\label{unitaryinv}
\end{equation}
Novel features of this RMT in addition to 
the conventional model (\ref{GE}) are:
\begin{itemize}
\item
Presence of the fermion determinant.
For small $m_f$, it expels Dirac eigenvalues of order $O(m_f)$
from the origin.
\item
Chiral structure of the random matrix.
Since $\{M, \gamma_5\}=0$ each eigenvalue $\lambda$ is accompanied by
its mirror image $-\lambda$. The Coulomb repulsion between
these pairs prevents them from populating around the origin.
\end{itemize}
Among the quantities calculable from this type of RMTs,
of particular interest are the distribution and correlation of
soft eigenvalues of the Dirac operator
as their accumulation is responsible for the
spontaneous breaking of the chiral symmetry via\refnote{\cite{BC}}
\begin{equation}
\Sigma=|\langle \overline{\psi}\psi \rangle| =
\frac{\pi\,\rho(0)}{N}.
\end{equation}
Due to the above two features, correlations of soft
eigenvalues are expected to deviate from the universal sine law 
(\ref{sine}). 
Verbaarschot and his collaborators\refnote{\cite{SV}} have calculated
correlations of eigenvalues of order $\lambda\sim O(1/N)$
from the gaussian ensemble
\begin{equation}
d\mu(W)=dW\,{\rm e}^{-N\Sigma\,{\rm tr}\,W^\dagger W},
\label{gaussmeasure}
\end{equation}
appealing to the conjecture of universality.
These analytic predictions have been compared first to the  
numerical data from the simulation of 
the instanton liquid model\refnote{\cite{V94}}, 
and more recently to that of 
the lattice gauge theory\refnote{\cite{V96, BMSVW}}.
The agreements are impressive.

In this article I shall review the proof of universality of the
microscopic spectral correlations of the model
(\ref{chGUE+m})
for $m_f=0$ and with single-trace potential measures\footnote{%
Microscopic universality 
for the gaussian potential plus certain terms which breaks 
the invariance (\ref{unitaryinv}) is proved in refs.\cite{JSV},
\cite{JNZ}.
See also ref.\cite{BHZ} for a perturbative treatment
of a quartic potential.}, 
in order to justify partly 
the use of gaussian ensemble (\ref{gaussmeasure}) and
to corroborate the above mentioned agreements.
Rather than to follow closely 
the original proof\refnote{\cite{Nis,ADMN1}},
I shall disguise it with the conventional approach of 
{\bf Q}, {\bf P} operators\refnote{\cite{DGZ}} 
and with the method employed in a recently proposed alternative 
proof\refnote{\cite{KF1,KF2}},
in a hope that borrowing notions from quantum mechanics
may add the proof some pedagogical flavor.
At the end of the article I shall address a problem
associated with multi-criticality.

\section{ORTHOGONAL POLYNOMIAL METHOD}

We start from recalling basic technology of random matrices:
the orthogonal polynomial method.
We consider the chiral unitary ensemble ($\chi$UE)
\begin{equation}
Z_{\chi\rm UE}=\int_{2N\times 2N\, 
\mbox{\scriptsize hermite}}\!\!\!\!\!\!\!\!\!\!
dM\,{\rm e}^{-N\,{\rm tr}\,V(M^2)}\, |{\rm det}\,M|^\alpha , 
\ \ \ \ \ \ 
M=
\left(
\begin{array}{cc}
0  &  W^\dagger \\
W & 0
\end{array}
\right)
\end{equation}
capturing the global symmetries of $N_c\geq 3$ QCD$_4$ 
with $N_f=\alpha$ massless fundamental fermions,
as well as the unitary ensemble (UE) without
chiral structure
\begin{equation}
Z_{\rm UE}=\int_{N\times N\, \mbox{\scriptsize hermite}}
\!\!\!\!\!\!\!\!\!\!
dM\,{\rm e}^{-N\,{\rm tr}\,V(M^2)}\, |{\rm det}\,M|^{\alpha}
\end{equation}
modeling QCD$_3$\refnote{\cite{VZ}}. 
We allow non-integer $\alpha >-1$, for they can be treated
on the same footing as integer cases.
Both ensembles allow eigenvalue representations\refnote{\cite{Mor}}
\begin{eqnarray}
Z_{\chi{\rm UE}}
&=&
\int_0^\infty \prod_{i=1}^N 
\left( d\lambda_i\,\lambda^{\alpha}_i\, 
{\rm e}^{-N V(\lambda_i) }\right)
\left| \det_{i j} \lambda^{i-1}_j \right|^2,\ \ \ \ \ \ 
(\lambda_i: \mbox{eigenvalues of }W^\dagger W)\\ 
Z_{{\rm UE}}
&=&
\int_{-\infty}^\infty 
\prod_{i=1}^N \left( d\lambda_i\,|\lambda_i|^{\alpha}\, 
{\rm e}^{-N V(\lambda_i^2) }\right)
\left| \det_{i j} \lambda^{i-1}_j \right|^2.\ \ \ 
(\lambda_i: \mbox{eigenvalues of }M)
\end{eqnarray}
Inside the integrals,
$\det \lambda^{i-1}_j$ may be replaced (up to an irrelevant constant)
with polynomials 
$\det P_{i-1}(\lambda_j)$
orthonormal with respect to the measure
$d\lambda\,|\lambda|^\alpha {\rm e}^{-N\,V(\lambda)}$:
\begin{equation}
\delta_{nm}=
\int d\lambda\,|\lambda|^\alpha {\rm e}^{-N\,V(\lambda)}\,
P_n(\lambda) \, P_m(\lambda)
=\int d\lambda\,\psi_n(\lambda) \, \psi_m(\lambda) .
\label{orthogonality}
\end{equation}
Here we have introduced
one-particle wave functions of a free fermion
(at the $n$-th state)
\begin{equation}
\psi_n(\lambda)=
|\lambda|^{\alpha/2} {\rm e}^{-N\,V/2}\,P_n(\lambda).
\end{equation}
The fermionic nature comes from the Vandermonde determinant.
The wave functions at first $N$ levels comprise the Dirac sea, i.e. 
the $N$-particle ground-state wave function
\begin{equation}
\Psi_N(\{\lambda\})=
\det_{1\leq n,m\leq N} \psi_{n-1}(\lambda_m).
\end{equation}
Using this ground-state wave function,
the vacuum expectation value of unitary invariant observables 
is written as
\begin{equation}
\left\langle {\cal O} \right\rangle=
\int d^N \lambda\,\Psi_N (\{\lambda\})\,{\cal O}\,\Psi_N (\{\lambda\}).
\end{equation}
Namely, spectral correlators are related to $\psi_n$ by
\begin{equation}
\rho_N(\lambda_1,\cdots,\lambda_p)\equiv
\left\langle \prod_{a=1}^p \frac{1}{N} 
{\rm tr}\, \delta (\lambda_a-M)\right\rangle
\stackrel{\mbox{\small Wick's theorem} \atop {\downarrow}}{=}
\det_{1\leq a,b \leq p} K_N(\lambda_a,\lambda_b)
\end{equation}
where $K_N$ is the projector to the Dirac sea:
\begin{eqnarray}
K_N(\lambda,\lambda')&=&
\sum_{n=0}^{N-1} \psi_n(\lambda)\psi_n(\lambda')\nonumber\\
&=&
q_N\,\frac{\psi_{N}(\lambda)\psi_{N-1}(\lambda')-
\psi_{N-1}(\lambda)\psi_{N}(\lambda')}{\lambda-\lambda'}.
\label{CD}
\end{eqnarray}
Use is made of Christoffel-Dalboux formula in the last line,
and $q_N$ is a constant defined immediately below.

We note that the orthogonality relation 
(\ref{orthogonality}) for the 
$\chi$UE case ($\int_0^\infty d\lambda$)
can absorbed into the UE case ($\int_{-\infty}^\infty d\lambda$) 
by the change of variables $\lambda\rightarrow\sqrt{\lambda}$
accompanied by the redefinition 
$\alpha\rightarrow \frac{\alpha-1}{2}$ 
(or $\alpha\rightarrow \frac{\alpha+1}{2}$) and
$P_{2n}(\sqrt{\lambda})\rightarrow P_n(\lambda)$
(or $\sqrt{\lambda}\,P_{2n+1}(\sqrt{\lambda})\rightarrow P_n(\lambda)$).
Thus in the following we need only to consider the UE case,
where orthogonal polynomials have definite parities.

Now let us for a while concentrate on $\alpha=0$,
where the logarithmic component of the potential is absent.
In this case one may represent the Heisenberg algebra
$[\frac{d}{d\lambda}, \lambda]=1$ 
by matrices {\bf Q} and {\bf P}
acting on the Hilbert space of orthonormal wave 
functions\refnote{\cite{DGZ}},
\begin{eqnarray}
\lambda\,\psi_n 
&=& q_{n+1}\psi_{n+1}+q_{n}\psi_{n-1}
\equiv\sum_m{{}\bf{Q}}_{nm}\,\psi_m,\\
\frac{d}{d\lambda}\,\psi_n 
&=&-\frac{N}{2} V'\,\psi_n + 
(\mbox{linear comb.~of }\psi_{n-1},\psi_{n-3},\cdots)
\equiv \sum_m{{}\bf{P}}_{nm}\,\psi_m.
\label{P}
\end{eqnarray}
(Anti-)self adjointness of multiplication and
differentiation operators inherit to the (anti-)symmetry
of the representing matrices,
\begin{eqnarray}
&
\int d\lambda\, \left(\lambda\,\psi_n \right) \psi_m=
\int d\lambda\, \psi_n \left(\lambda\,\psi_m \right)
&\Rightarrow 
{{}\bf{Q}}_{nm}={{}\bf{Q}}_{mn},\\
&
\int d\lambda\, \left(\frac{d}{d\lambda} \psi_n \right) \psi_m=
-\int d\lambda\, \psi_n \left(\frac{d}{d\lambda} \psi_m \right)
&\Rightarrow
{{}\bf{P}}_{nm}=-{{}\bf{P}}_{mn}.
\label{antisym}
\end{eqnarray}
Although eq.(\ref{P}) and the antisymmetry (\ref{antisym})
are enough to determine the {\bf P} matrix as
\begin{equation}
{{}\bf{P}}_{nm}=
\mp\frac{N}{2}\,V'({{}\bf{Q}})_{nm} 
\ \ ( n \vcenter{\hbox{$<$}\hbox{$>$}} m ),
\label{explicitP}
\end{equation}
we shall not use this fact except for fixing a 
integration constant in the sequel.

At this stage we consider the situation when the coefficient $q_n$
become single-valued
in the limit $n, m\simeq N \gg 1$,
\begin{equation}
q_n\sim q_{n\pm1}\sim q=O(1),
\end{equation}
a crucial assumption which is violated in matrix ensembles
with multi-band spectra\refnote{\cite{CMM}}.
Then the matrix elements of 
${{}\bf{Q}}$, ${{}\bf{P}}$ for $n, m =N + O(1)$ 
take the forms:
\begin{equation}
\!\!
{{}\bf{Q}}_{nm}=
{\left(
\begin{array}{cccccc}
\ddots & \ddots &     &    &   &         \\
\ddots & \ddots & \ddots &    &   &         \\
    & \ddots & 0   & q  &   &          \\
    &     & q   & 0  & q &          \\
    &     &     & q  & 0 & q        \\
    &     &     &    & q & \ddots  
\end{array}
\right)},\ \ 
{{}\bf{P}}_{nm}=N
{\left(
\begin{array}{cccccc}
\ddots & \ddots & \ddots & \ddots  & 0    & \ddots    \\
\ddots & \ddots & \ddots & 0    & -p_2 & 0      \\
\ddots & \ddots & 0   & -p_1 & 0    & -p_2   \\
\ddots & \ddots & p_1 & 0    & -p_1 & 0     \\
\ddots & p_2 & 0   & p_1  & 0    & -p_1  \\
\ddots & 0   & p_2 & 0    & p_1  & \ddots  
\end{array}
\right)}.
\label{Q}
\end{equation}
A matrix of the above form of ${{}\bf{P}}$ can always be written as
\begin{equation}
{{}\bf{P}}=
\frac{N}{2} 
\sum_{k\geq 0} a_{k}\,{{}\bf{Q}}^{2k} {{}\bf{\Lambda}}
\equiv \frac{N}{2} {A\left({{}\bf{Q}}\right)} {{}\bf{\Lambda}},
\label{PQ}
\end{equation}
where the $\Lambda$ matrix stands for the antisymmetrizer
\begin{equation}
{{}\bf{\Lambda}}_{nm}=
\left(
\begin{array}{cccccc}
\ddots & \ddots &     &    &    &          \\
\ddots & \ddots & \ddots &    &    &          \\
    & \ddots & 0   & -1 &    &          \\
    &     & 1   & 0  & -1 &          \\
    &     &     & 1  & 0  & -1       \\
    &     &     &    & 1  & \ddots 
\end{array}
\right).
\end{equation}
When $\alpha\neq 0$, $\frac{d}{d\lambda}$ can not be represented
on the space of orthogonal polynomials as it involves 
a term proportional to ${1}/{\lambda}$.
However one may verify\refnote{\cite{KF2}} that the operator
$\frac{d}{d\lambda}-\frac{(-)^{\hat{n}}\alpha}{2\lambda}$ is indeed
represented by the same {\bf P} matrix as in (\ref{PQ}).
Here $\hat{n}$ is the number operator 
$\hat{n}\, \psi_n = n\, \psi_n$, 
and the existence of the signature $(-)^{\hat{n}}$ 
preserves the antisymmetry of {\bf P},
\begin{equation}
\psi_{2m+1} \left( (-)^{\hat{n}} \psi_{2n}\right) = 
-\left( (-)^{\hat{n}} \psi_{2m+1}\right)\psi_{2n}.
\end{equation}
In the next section eqs.(\ref{Q}) and (\ref{PQ}) are utilized
to prove the microscopic universality.

\section{PROOF OF UNIVERSALITY}

Here we re-exhibit the asymptotic forms of
recursion relations (\ref{Q}), (\ref{PQ}) 
satisfied by the orthonormal wave functions ($\alpha=0$):
\begin{eqnarray}
\lambda\,\psi_n &=& 
q \left(\psi_{n+1}+\psi_{n-1} \right),\\
\frac{d}{d\lambda}\,\psi_n &=& 
\frac{N A(\lambda)}{2} \left(-\psi_{n+1}+\psi_{n-1} \right).
\end{eqnarray}
They are combined to yield unharmonic analogues of 
lowering and raising operators of a harmonic oscillator
at highly excited levels $n\gg 1$:
\begin{eqnarray}
\left(
\frac{1}{N A(\lambda)} \frac{d}{d\lambda}
+\frac{\lambda}{2q}
\right)
\psi_n &=& \psi_{n-1},
\\
\left(
-\frac{1}{N A(\lambda)} \frac{d}{d\lambda}
+\frac{\lambda}{2q}\right)
\psi_n &=& \psi_{n+1}.
\end{eqnarray}
The requirement that two processes
$\psi_n {
\nearrow\psi_{n+1}\searrow 
\atop
\searrow\psi_{n-1}\nearrow 
} \psi_{n}$
must commute amounts to
the suppression of the commutator of the 
raising and lowering operators in the large-$n$ and $N$ limit, 
\begin{equation}
\left[\frac{\lambda}{2q}\ ,\  
\frac{1}{N A(\lambda)} \frac{d}{d\lambda}
\right] = O (\frac{1}{N^{*}})\ll 1 .
\end{equation}
Thus we are lead to 
a second-order differential equation
for $\psi_n$,
\begin{equation}
\left[
\left( \frac{1}{N A(\lambda)} \frac{d}{d\lambda}\right)^2 
+1-\left(\frac{\lambda}{2q}\right)^2 \right]
\psi_n = 0.
\end{equation}

As mentioned in the end of the last section,
the only modification for $\alpha\neq 0$ case is
the replacement
\begin{equation}
\frac{d}{d\lambda} \psi_n 
\rightarrow 
\left(\frac{d}{d\lambda} -\frac{(-)^n\alpha}{2\lambda} \right)\psi_n.
\end{equation}
The resulting equation takes the form
\begin{equation}
\left[\frac{1}{N^2}
\frac{1}{A(\lambda)} \left(
\frac{d}{d\lambda} 
+\frac{(-)^n\alpha}{2\lambda} \right)
\frac{1}{A(\lambda)} \left( 
\frac{d}{d\lambda} 
-\frac{(-)^n\alpha}{2\lambda} \right)+
1-\left(\frac{\lambda}{2q}\right)^2 \right]
\psi_n = 0.
\label{diffeq}
\end{equation}

Let us first take the macroscopic (large-$N$) limit:
\begin{equation}
N\rightarrow\infty,\, \  \lambda:\mbox{ fixed}
\end{equation}
of (\ref{diffeq}).
Since $1/N$ plays the role of $\hbar$ in this limit
and thus eq.(\ref{diffeq}) is reduced to 
\begin{equation}
\left[\frac{1}{N^2}
\frac{1}{A(\lambda)^2} 
\frac{d^2}{d\lambda^2}
+
1-\left(\frac{\lambda}{2q}\right)^2 \right]
\psi_n = 0,
\end{equation}
we obtain the WKB solution\refnote{\cite{BZ}} \footnote{
This process was initiated in the context of stochastic quantization
of matrix models\refnote{\cite{AK}}, although their Fokker-Planck
equation differs from eq.(\ref{diffeq}).
I thank C.~F.~Kristjansen for remarks on this point.
}:
\begin{equation}
\psi_n(\lambda)\propto {\cos}
\left( N\int_0^\lambda d\lambda\,A(\lambda)
\sqrt{1-\left(\frac{\lambda}{2q}\right)^2}
+\frac{n\pi}{2}
\right).
\end{equation}
Substituting this solution to (\ref{CD}), we obtain
the kernel in the large-$N$ limit:
\begin{eqnarray}
K(\lambda,\lambda')
\equiv\lim_{N\rightarrow\infty}K_N(\lambda,\lambda')
&\propto& 
\frac{\sin\left( 
N\int_{\lambda'}^{\lambda} d\lambda\,A(\lambda)
\sqrt{1-\left(\frac{\lambda}{2q}\right)^2}\right)}{\lambda-\lambda'}
,\\
\rho(\lambda)=K(\lambda,\lambda)&=&{\rm const.}
A(\lambda)\sqrt{1-\left(\frac{\lambda}{2q}\right)^2}.
\label{rhoA}
\end{eqnarray}
The constant is not fixed within this approach, though
it can fixed to be $1/\pi$ if we use the explicit form of
the {\bf P} operator (\ref{explicitP}).
The meanings of $q$ and $A(\lambda)$ become clear at this stage:
$\pm 2q$ stand for the edges of the spectrum and
$A(\lambda)$ the deviation of the spectral envelope from
Wigner's semi-circle.

Now we proceed to take the microscopic limit: 
\begin{equation}
N\rightarrow\infty,\ \lambda\rightarrow 0,\ 
z\equiv N\lambda:\ \mbox{fixed}.
\end{equation}
We have already assumed the single-valuedness of
the recursion coefficients $q_n$, which is known to be
the sufficient condition for $A(0)=\pi \rho(0)>0$. (In other words,
we have assumed the chiral symmetry breaking as an input.)
Then eq.(\ref{diffeq}) is reduced to
\begin{equation}
\left[
\frac{1}{A(0)^2} 
\left( \frac{d}{dz}+\frac{(-)^n\alpha}{2z} \right)
\left( \frac{d}{dz}-\frac{(-)^n\alpha}{2z} \right)
+1\right]
\psi_n = 0.
\label{besseleq}
\end{equation}
Its solution which is regular at $z=0$ is a Bessel function:
\begin{equation}
\psi_n\propto\sqrt{z}\,J_{\frac{\alpha\mp 1}{2}}\left(A(0)z\right).
\ \ \ \ \left( n={\mbox{\small even} \atop \mbox{\small odd}} \right)
\end{equation}
Within either even or odd sector of the wave functions,
the $n$-dependence enters only through
$A_n(0)$. Thus we have
\begin{equation}
\psi_n-\psi_{n-2}
\sim
\frac{d}{dA(0)}\psi_n
\propto
z^{\frac32}\left(
J_{\frac{\alpha\mp3}{2}}\left(A(0)z\right)-
J_{\frac{\alpha\pm1}{2}}\left(A(0)z\right) \right)
.
\ \ \ \ \left( n={\mbox{\small even} \atop \mbox{\small odd}} \right)
\end{equation}
Substituting the above solutions again into (\ref{CD}),
we obtain universal forms of the microscopic kernels 
(called the Bessel kernels\refnote{\cite{NS}}) 
\begin{equation}
K_s(z,z')\equiv
\lim_{N\rightarrow\infty}\frac1N K_N(\frac{z}{N},\frac{z'}{N}),
\end{equation}
\begin{eqnarray}
\chi{\rm UE}:&& 
K_{s} (z, z')=
\pi \rho(0) \sqrt{z\,z'}
\frac{
z J_{\alpha+1} (2\pi \rho(0) z) 
  J_\alpha (2\pi \rho(0) z' )-
(z\leftrightarrow z')
}{z^2-{z'}^2}, 
\label{chUEkernel}\\
{\rm UE}:&& 
K_{s} (z, z')=
\frac{\pi \rho(0)}{2}\sqrt{z\,z'} 
\frac{
J_{\frac{\alpha+1}{2}}(\pi \rho(0) z)
J_{\frac{\alpha-1}{2}}(\pi \rho(0) z')-
(z\leftrightarrow z')
}{z-z'} .
\label{UEkernel}
\end{eqnarray}
and microscopic spectral densities
$\rho_s(z)=K_s(z,z)$,
\begin{eqnarray}
\chi{\rm UE}:&& 
\rho_{s} (z)=
\left( \pi \rho(0) \right)^2
|z|
\left(
J_\alpha^2 -
J_{\alpha+1}
J_{\alpha-1}
\right)(2\pi\rho(0)  z), 
\label{bessel}\\
{\rm UE}:&& 
\rho_{s} (z)=\left(\frac{\pi\rho(0)}{2}\right)^2 z \left(  
J_{\frac{\alpha+1}{2}}^2 + J_{\frac{\alpha-1}{2}
}^2      
- J_{\frac{\alpha+1}{2}}J_{
\frac{\alpha-3}{2}}   
- J_{\frac{\alpha-1}{2}}J_{\frac{\alpha+3}{2}}   
\right) \left( {\pi\rho(0) z} \right) .
\end{eqnarray}
In the above, the integration constants are fixed by requiring
$\lim_{z\rightarrow \infty}\rho_s(z) = \rho(0)$.
When measured in the unit of mean level spacing $1/{\rho(0)}$, 
$K_s(z, z')\,dz$ contains no free parameter.
After this rescaling,
the Bessel kernels (\ref{chUEkernel}) and (\ref{UEkernel})
approach the sine kernel (\ref{sine}) 
in the limit $z,\,z'\rightarrow \infty$,
$z-z' = O(1)$ as the repulsive effects due to the fermion determinant
and the chiral structure become negligible.

\section{DISCUSSIONS}

In this article I have presented a proof of universality of
microscopic correlations for RMT modeling QCD$_{3, 4}$
coupled to massless quarks,
within a single-trace potential class.
The spectral correlation function is shown to be insensitive to the
choice of matrix measures as long as the macroscopic
spectrum is supported on a single interval.

Universality of the Bessel kernel
is valid even beyond large-$N$ matrix models.
Namely, the zero-dimensional reduction of the SU$(N_f)$ 
$\sigma$-model of pions also yields sum rules\refnote{\cite{LS}}
which are identical to the moments of (\ref{bessel}).
Together with the numerical agreements mentioned in the introduction,
it is very convincing that this wide range of universality
encompasses QCD$_4$. 

Recently it also became possible to incorpolate quarks
whose masses are within the same microscopic range
as the Dirac eigenvalues\refnote{\cite{DN1,DN2}}. 
We expect that the exactness of those universal 
correlations be verified by numerical simulations
in a foreseeable future.

On the other hand, one may address a question: whether
RMT can describe chiral symmetric phases of QCD
such as Higgs phase or the large $N_f$ case.
In these cases, accumulation of the soft eigenvalues 
is not strong enough to form a chiral condensate, 
although the theory is still interacting.
There is no satisfactory answer to it so far,
yet one may check the existence of universality
within RMT. 
If the microscopic universality in the vicinity of the origin
is broken for random matrix ensembles with 
$\rho(0)\propto|\langle \overline{\psi}\psi \rangle| =0$,
we may not expect to extract informations from RMT.
In view of the above, we are lead to analyze a RMT with a double-well
potential such as
\begin{equation}
Z = \int_{N\times N\ {\rm matrices}}\!\!\!\!\!\!\!\!\!\!
dM\, {\rm e}^{-N\,{\rm tr}\,(-M^2+g M^4)},\ \ g=\frac14
\label{crit}
\end{equation}
whose coupling is tuned so as the macroscopic spectrum
$\rho(\lambda)$ to have a double zero at $\lambda=0$
\refnote{\cite{Shi}}.
According to (\ref{rhoA}) it corresponds to $A(\lambda)$
of the form
\begin{equation}
A(\lambda)\sim \lambda^2 + O(\lambda^4).
\end{equation}
Then eq.(\ref{diffeq}) suggests a possible scaling limit:
\begin{equation}
N\rightarrow\infty,\ \lambda\rightarrow 0,\ 
z\equiv N^{1/3}\lambda:\ \mbox{fixed}.
\label{m=1}
\end{equation}

However, the whole procedure must be reconsidered 
since the recursion coefficients $q_n$ become
double-valued in the large-$n$ and $N$ limit.
A working hypothesis is to assume the following asymptotic 
form\refnote{\cite{CM}}
\begin{equation}
q_n = q_c + N^{-1/3} (-)^n\, f(t)
+ N^{-2/3} g(t) + O(N^{-1}),\ \ \ 
t\equiv N^{2/3}\left(1-\frac{n}{N}\right) .
\end{equation}
Then the function $A(\lambda)$ in the limit (\ref{m=1})  
is modified by a constant term
involving $f'(0)$ and $g(0)$, which are the same order
$O(N^{-2/3})$ as $\lambda^2$ under the limit (\ref{m=1}),
\begin{equation}
A(\lambda)=q_n (-2q_c^2 + q_n^2 + q_{n+1}^2 + \lambda^2)
\sim N^{-2/3}\, (2g(0)-(-)^n f'(0) + z^2).
\end{equation}
The numerical values of $f'(0)$ and $g(0)$ are
determined by solving Painlev\'{e} II 
equation (the microscopic limit of the Heisenberg algebra
$[{\bf Q} , {\bf P} ]=1$) under an appropriate
boundary condition\refnote{\cite{CM}}. 
We may also have $g$ run toward $g_c=1/4$ with
\begin{equation}
\gamma=N^{2/3} (g-g_c):\ \mbox{fixed}
\end{equation}
while retaining the macroscopic spectrum intact.
In this case the functions $f'(t)$ and $g(t)$ ought to be
evaluated at $t=\gamma$ instead of $t=0$.
These numerical values enter 
the resulting second order linear differential equation 
for the wave function $\psi_n$ (the counterpart of (\ref{besseleq})).
Thus the macroscopic spectral density $\rho(\lambda)$ is not 
sufficient to determine
the microscopic correlation, but the latter depends upon the way how to
approach the critical point.
Microscopic universality holds in a weak sense for these cases, 
for Painlev\'{e} II equation is universally derived from a class 
of the critical potentials.
Details of the analysis will appear 
elsewhere\refnote{\cite{ADMN2}}.

\section{ACKNOWLEDGEMENTS}
I thank M.~Prasza\l owicz
and the organizers of the Workshop for their hospitality.
Thanks are also due to P.~H.~Damgaard 
for collaboration and discussions summarized
in this article. 
The work of SMN is supported in part by 
JSPS Postdoctoral Fellowships for Research Abroad, and
by Nishina Memorial Foundation.

\end{document}